# Adaptation to the Primary User CSI in Cognitive Radio Sensing and Access

Yuan Lu, Alexandra Duel-Hallen

Department of Electrical and Computer Engineering
North Carolina State University
Raleigh, NC, 27606
{ylu8, sasha}@ncsu.edu

*Abstract*—In Cognitive Radio (CR) networks, multiple secondary network users (SUs) attempt to communicate over wide potential spectrum without causing significant interference to the Primary Users (PUs). A spectrum sensing algorithm is a critical component of any sensing strategy. Performance of conventional spectrum detection methods is severely limited when the average SNR of the fading channel between the PU transmitter and the SU sensor is low. Cooperative sensing and advanced detection techniques only partially remedy this problem. A key limitation of conventional approaches is that the sensing threshold is determined from the miss detection rate averaged over the fading distribution. In this paper, the threshold is adapted to the instantaneous PU-to-SU Channel State Information (CSI) under the prescribed collision probability constraint, and a novel sensing strategy design is proposed for overlay CR network where the instantaneous false alarm probability is incorporated into the belief update and the reward computation. It is demonstrated that the proposed sensing approach improves SU confidence, randomizes sensing decisions, and significantly improves SU network throughput while satisfying the collision probability constraint to the PUs in the low average PU-to-SU SNR region. Moreover, the proposed adaptive sensing strategy is robust to mismatched and correlated fading CSI and improves significantly on conventional cooperative sensing techniques. Finally, joint adaptation to PU channel gain and SU link CSI is explored to further improve CR throughput and reduce SU collisions.

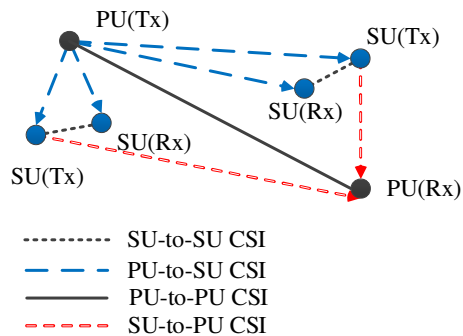

Fig. 1: Types of CSI in a typical CR Scenario.

## I. Introduction

Cognitive radio (CR) is an emerging technology that can potentially improve spectrum utilization and has drawn broad interest from researchers in recent years. In CR networks, secondary users (SUs) attempt to communicate over a set of channels without severely limiting activities of the primary users (PUs). The SUs employ a sensing strategy, or a medium access control (MAC) algorithm, to make sensing and access decisions. A spectrum sensing method is an integral component of any sensing strategy. Classical sensing approaches include matched filtering and energy detection [1]. To provide sufficient protection to the PU receivers, a constant detection rate (CDR) of the PU signals [2] is required. To avoid the "hidden node" problem and to protect the PUs, it is desirable to maintain sensing accuracy even when the signal from the PU transmitter to the SU detector is weak (low PU-to-SU SNR) [3]. However, individual SU sensing decisions become unreliable for fading PU-to-SU channels with low average SNR. To remedy this problem, cooperative spectrum sensing approaches [4] and feature-based sensing [1] were proposed, but the gain of these approaches is limited for realistic CR networks.

*Related work:* In [5], PU-to-SU channel gain (see Fig. 1) is employed as a criterion in choosing channels for sensing. However, uniform PU activity across a wide band of channels was assumed. Moreover, a fixed threshold based on the false alarm rate constraint was employed although in practice the threshold should be chosen to satisfy a miss detection rate constraint [2]. Sensing threshold adaptation for single channel CR networks was investigated based on SU transmission power [6], the Channel State Information (CSI) between SU pairs [7], the amount of interference caused to PUs in case of missed detection [8], and the sensed SNR [9]. However, threshold adaptation has not been incorporated into sensing strategy design for multichannel CR networks.

*Contribution*: We design a sensing strategy for overlay CR networks that adapts to the instantaneous SNR of the signal between the PU transmitter and the SU sensor, i.e. PU-to-SU CSI illustrated in Fig. 1. To the best of our knowledge, only [9] has explored such threshold adaptation. However, in [9], only one PU pair, one SU pair, and one channel were assumed, and unrealistic constraints that require the knowledge of the PU-to-PU SNR statistics and the instantaneous channel gain between the SU transmitter and the PU receiver at the SU (SU-to-PU CSI in Fig. 1) was employed. We consider multiple SU pairs that compete for available channels under the hardware constraints. To offer sufficient protection to the primary network, we impose a constraint on the instantaneous miss detection probability at each SU. The resulting instantaneous false alarm probability is incorporated into the belief update and reward computation of the sensing strategy. By selecting to sense channels with high instantaneous PU-to-SU SNR, the

This research was supported by the NSF grant CNS-1018447.

proposed policy reduces false alarm rate, improves sensing decisions, and increases the CR network throughput.

Since the PU range is often much larger that the SU range, SUs converge to similar sensing decisions and suffer from network congestion when the myopic, or greedy, strategy is used [10]. The proposed adaptive sensing strategy randomizes sensing decisions of different SU detectors and helps to resolve SU collisions since the received channel gain from the PU transmitter varies over SU locations and frequencies. Thus, the proposed detection method converts the conventional myopic strategy into a randomized sensing strategy. Moreover, we combine the proposed sensing threshold adaptation with the channel-aware myopic sensing strategy in [11] that adapts the reward to the CSI of the SU link.

We also investigate practical feasibility of adaptive sensing threshold control. First, this policy requires the knowledge of the PU-to-SU channel gain prior to sensing, which can be obtained directly from a channel gain map [12] if available. Otherwise, such information can be acquired from previous spectrum sensing or during the "silence" phase [13] when an SU does not have data to transmit and/or has sensed a channel that is occupied by a PU. This CSI is likely to be noisy and outdated, will require estimation and prediction, and CSI mismatch at the sensor is likely. To maintain the miss detection rate constraint, we incorporate the CSI error into the sensing strategy design and investigate robustness to CSI mismatch. Second, we validate performance of the proposed strategy for multipath and correlated shadow fading channel models. Finally, adaptive threshold control is compared with cooperative sensing detection for realistic network scenarios.

The rest of this paper is organized as follows. In section II, we formulate the problem and discuss sensing threshold adaptation. Myopic PU-to-SU CSI-aided sensing strategy is described and combined with reward adaptation to the CSI of the SU link in section III. Numerical results are presented in section IV. Finally, we draw conclusions in section V.

## II. ADAPTIVE SENSING THRESHOLD CONTROL.

In this paper, we consider an overlay CR network [1] with $M$ SU transmitter-receiver pairs and $N$ orthogonal channels. The SUs can only access spectrum when active PUs are not detected in the neighborhood and are required to sense the spectrum before accessing any channel. All SUs make their own sensing and access decisions autonomously without the coordination of a central controller.

Suppose $\mathbf{y}^{mn}(t) = [y_1^{mn}(t),...,y_\nu^{mn}(t)]$ is the signal received by the sensor of the $m^{th}$ SU on the $n^{th}$ channel at the time slot $t$, where $\nu$ is the number of collected samples. The components $y_i^{mn}(t)$ contain independent and identically distributed (i.i.d) Gaussian noise terms with unit variance. If the PU transmitter is active, they also contain the PU signal. The instantaneous PU-to-SU SNR per sample $\lambda^{mn}$ has the distribution $f_{\lambda^{mn}}(\lambda^{mn})$. Assume energy detection [4]. If the PU signal is not present during the sensing period (null hypothesis $H_0$), the output decision statistic $S(\mathbf{y}^{mn}(t))$ follows a central Chi-square distribution with $2\nu$ degrees of freedom. If the PU signal is present (alternative hypothesis $H_1$), $S(\mathbf{y}^{mn}(t))$ follows a non-central Chi-square distribution with $2\nu$ degrees of freedom and a non-centrality parameter $2\nu\lambda^{mn}(t)$. If the decision statistic $S(\mathbf{y}^{mn}(t))$ is larger than the detection threshold $\tau^{mn}(t)$, the spectrum sensor accepts the alternative hypothesis $H_1$ and vice versa.

The instantaneous miss detection probability $p_{MD}^{mn}(t)$ and the false alarm probability $p_{FA}^{mn}(t)$ are given by

$$p_{MD}^{mn}(t) = \Pr[\mathbf{y}^{mn}(t) < \tau^{mn}(t) | H_1] \\ = 1 - Q_\nu(\sqrt{2\nu\lambda^{mn}(t)}, \sqrt{\tau^{mn}(t)}), \quad (1)$$

and

$$p_{FA}^{mn}(t) = \Pr[\mathbf{y}^{mn}(t) > \tau^{mn}(t) | H_0] \\ = \frac{\Gamma(\nu, \tau^{mn}(t)/2)}{\Gamma(\nu)}, \quad (2)$$

where $Q_\nu(\cdot,\cdot)$ is the generalized Marcum Q-function, $\Gamma(\cdot,\cdot)$ and $\Gamma(\cdot)$ are the upper incomplete gamma function and the complete gamma function, respectively.

Conventionally the threshold is fixed, so $\tau^{mn}(t) = \tau^{mn}$. In this case the probability of miss detection is given by the expectation

$$p_{MD}^{mn} = 1 - \int_{\lambda^{mn}} Q_\nu(\sqrt{2\nu\lambda^{mn}}, \sqrt{\tau^{mn}}) f_{\lambda^{mn}}(\lambda^{mn}) d\lambda^{mn}, \quad (3)$$

and the probability of false alarm is

$$p_{FA}^{mn} = \frac{\Gamma(\nu, \tau^{mn}/2)}{\Gamma(\nu)}. \quad (4)$$

We propose to adjust the threshold according to the instantaneous PU-to-SU CSI. Assuming the ideal CSI knowledge, the detector employs *the instantaneous false alarm and miss detection probabilities* (1,2) instead of averaging these probabilities over the fading distribution.

Both the conventional and the adaptive detectors must satisfy the miss detection rate constraint $p_{MD,Target}$. The detection threshold is computed by inverting the miss detection probability:

$$\tau^{mn}(t) = p_{MD}^{mn\ -1}(p_{MD,Target}), \quad (5)$$

where $p_{MD}^{mn}$ is given by (3) for the traditional energy detector and by $p_{MD}^{mn}(t)$ in (1) for the adaptive threshold selection.

Since the range of the PUs is usually much larger than the range of the SUs, all SUs in the neighborhood have similar PU SNR statistics. Thus, in the fixed threshold case, the false alarm probability and the threshold are time-invariant and are likely to take on the similar values for neighboring SUs. However, for the proposed method, these parameters are time-variant and *will have different values across the CR spectrum for different SUs* due to spatial and frequency diversity in fading scenarios. Moreover, from (1,2) and (5), $p_{FA}(t)$ *decreases with* $\lambda^{mn}(t)$ *given* $p_{MD,Target}$. Thus, as the received power at the SU sensor increases, that SU can raise the detection threshold while maintaining a certain collision probability constraint. As a result, abrupt fluctuations in the received power caused by noise or interference will not be misidentified as PU signals, resulting in fewer wasted spectrum opportunities relative to the conventional detector case.

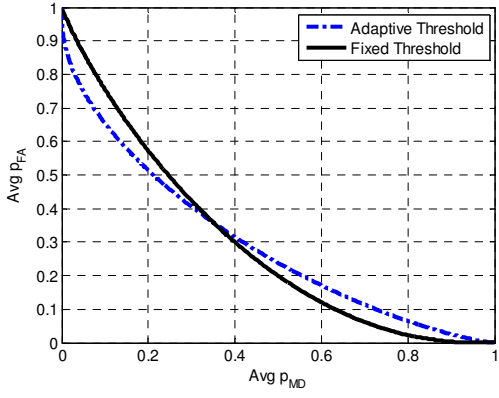

Fig. 2: Comparison of ROC curves for energy detection with fixed vs. adaptive threshold selection; Rayleigh fading; $\bar{\lambda} = -10 dB$; $\nu = 100$.

The receiver operating characteristic (ROC) curves of the energy detector with fixed and adaptive threshold selection under Rayleigh fading are compared in Fig. 2. The average PU-to-SU SNR $\bar{\lambda} = -10 dB$. When the prescribed probability of miss detection is under 0.3, computing the threshold adaptively provides a lower average false alarm probability and thus *improves sensing reliability*.

## III. MYOPIC SENSING STRATEGIES WITH ADAPTIVE THRESHOLD SELECTION

The PU traffic is modeled as a stationary Markov process with known transition probabilities evolving independently on each channel. For channel $n$ at the $m^{th}$ SU location, $p_{ij}^{mn}$ denotes the probability of transition from state $i$ to state $j$, where $i, j \in \{0(\text{busy}), 1(\text{idle})\}$. All PUs and SUs share the same slotted structure and are perfectly synchronized [10]. We assume that each SU can sense and then access only one channel at each time slot due to the hardware constraints. The belief vector $\boldsymbol{\theta}^m(t) = [\theta^{m1}(t),...,\theta^{mn}(t),...,\theta^{mN}(t)]$ is employed by the SUs to infer the current state of the PU traffic, where $\theta^{mn}(t)$ is the conditional probability that channel $n$ is available at time $t$ for the $m^{th}$ SU pair based on past sensing history [14]. The sensing result $a^m(t) = 1$ if a spectrum opportunity is correctly detected or if a missed detection occurs, and $a^m(t) = 0$ if a PU activity is correctly detected or when a false alarm occurs.

In this paper we consider *myopic*, or greedy, sensing policies where each SU makes sensing decisions selfishly without taking into account possible collisions with other CR users. Suppose the reward for SU $m$ on channel $n$ is $R^{mn}(t)$. At the first time slot $t = 1$, the initial belief vector is given by the stationary probabilities of the Markov process. Then at each time slot $t > 1$, SU $m$ chooses to sense the channel $n_*^m(t)$ by maximizing *the expected reward* $E[R^{mn}(t)]$:

$$n_*^m(t) = \arg\max_n \theta^{mn}(t) R^{mn}(t). \quad (6)$$

In the equations below, the false alarm and the probability of miss detection terms are given by (1,2) for the proposed adaptive threshold and (3,4) for the conventional fixed threshold energy detection, respectively. After sensing, the belief is corrected by the reliability of the spectrum sensor [15],

$\forall n = n_*^m(t), m = 1,...M,$

$$\theta_r^{mn}(t) = \begin{cases} \dfrac{(1-p_{FA}^{mn})\theta^{mn}(t)}{(1-p_{FA}^{mn})\theta^{mn}(t) + p_{MD}^{mn}(1-\theta^{mn}(t))}, & a^m(t) = 1 \\ \dfrac{p_{FA}^{mn}\theta^{mn}(t)}{p_{FA}^{mn}\theta^{mn}(t) + (1-p_{MD}^{mn})(1-\theta^{mn}(t))}, & a^m(t) = 0 \end{cases} \quad (7)$$

and then updated according to the Markov chain,

$$\theta^{mn}(t+1) = \begin{cases} p_{11}^{mn}\theta_r^{mn}(t) + p_{01}^{mn}(1-\theta_r^{mn}(t)), & \text{if } n_m^*(t) = n \\ p_{11}^{mn}\theta^{mn}(t) + p_{01}^{mn}(1-\theta^{mn}(t)), & \text{if } n_m^*(t) \neq n \end{cases} \quad (8)$$

where the process is repeated over the time horizon $t \in [1,T]$. When the instantaneous reliability parameter $p_{FA}^{mn}(t)$ is employed in (7,8) instead of the average $p_{FA}^{mn}$, more accurate estimation of the current PU traffic states results.

Finally, the reward is modified by the instantaneous probability of false alarm in the proposed policy. Suppose the reward for a fixed threshold strategy is given by $R_{FT}^{mn}(t)$. The corresponding sensing strategy with adaptive threshold update employs the reward

$$R_{AT}^{mn}(t) = (1 - p_{FA}^{mn}(t))R_{FT}^{mn}(t). \quad (9)$$

Thus, $R^{mn}(t)$ in (6) is given by $R_{FT}^{mn}(t)$ when the conventional sensing method is employed and by (9) for adaptive threshold selection.

By taking into account the sensing reliability when selecting channels to sense, *SUs will favor stronger PU-to-SU channels* since $p_{FA}(t)$ decreases with $\lambda^{mn}(t)$. This approach increases SU confidence relative to the conventional sensing method where the individual SU throughput is sacrificed to protect the PUs. Moreover, due to geographical separation that provides spatial and frequency diversity, SUs perceive distinct sensing reliabilities on each channel, resulting in different sensing decisions. *Thus, the proposed policy randomizes sensing decisions and reduces SU congestion*.

Conventionally the reward is given by the channel bandwidth, i.e.,

$$R_{FT}^{mn} = B^n. \quad (10)$$

However, when this reward is employed in the myopic policy, it results in severe CR network congestion and poor throughput. To reduce congestion, several strategies in the literature, e.g. [14], randomize sensing decisions or use negotiation while retaining the reward given by the channel bandwidth. However, the gains of these strategies are limited.

In [11], we proposed *to adapt the reward to the maximum achievable rate of the SU link*, i.e.,

$$R_{\text{FT}}^{mn} = C^{mn}(t) = B_n \log_2(1+\gamma^{mn}(t)), \quad (11)$$

where $\gamma^{mn}(t)$ is the instantaneous SNR of the $m^{th}$ SU pair on the $n^{th}$ channel, and $C^{mn}(t)$ is the channel capacity. This sensing strategy exploits spatial and frequency diversity, randomizes sensing decisions, and boosts the network throughput. It significantly outperforms other randomized strategies even when they employ adaptive transmission. This gain is due to adaptiation to the SU link CSI *prior to sensing*. We showed that this approach is robust to CSI mismatch and fading correlation and retains its gain when the reward is computed using realistic adaptive modulation [11].

In practice, sensing errors significantly degrade the throughput of all strategies in the literature under a realistic collision probability constraint, especially in the low PU-to-SU SNR region [3]. To remedy this problem, we can employ adaptive threshold control. These two types of adaptation, i.e. adaptation to PU-to-SU and SU-to-SU CSI, are tested individually and jointly in the numerical results below. Thus, *we evaluate the benefits of adaptive threshold control for both conventional and channel-aware myopic strategies and the gain of combined adaptation to the PU-to-SU and the SU-to-SU link CSI.*

## IV. NUMERICAL RESULTS

Consider a CR network with $M=20$ SU pairs and $N=40$ channels with the same bandwidth $B=1$. The transition probabilities of the PU traffic on all channels at all SU locations are $[p_{01}\ p_{11}] = [0.2\ 0.8]$. All SU-to-SU, PU-to-SU, and PU-to-PU channels are subject to independent Rayleigh fading unless stated otherwise. All SU-to-SU links are identically distributed on all channels with the average SNR $\bar{\gamma}$. Similarly, at all SU sensors the average PU signal SNR $\bar{\lambda}$ is the same on all channels. In this paper we focus on low average SNR from the PU transmitter to the SU sensor (PU-to-SU SNR in Fig. 1). Note that the PU receiver can be closer to the sensor than the PU transmitter, so the interference to the PU network (SU-to-PU SNR) can still be significant. We assume an overlay scenario where a miss detection results in a collision between the SU and the PU transmissions.

We employ a MAC scheme similar to [16] where an SU will transmit over a channel if it is sensed idle or go to sleep during the current time slot if it is sensed busy. If multiple SU pairs choose to sense the same channel and if that channel is idle, only one of them can transmit successfully. Moreover, we assume that SUs always have data to transmit. Finally, *the SU network throughput for any sensing strategy in this paper is computed under the assumption that adaptive transmission is employed after sensing with the accumulated reward given by the channel capacity.*

Since the generalized Marcum Q-function in (1) and its inverse in (5) are very computationally complex, we employ the Gaussian approximation that holds for $v \gg 1$ [1].

### A. Throughput gain of PU-to-SU CSI Adaptation

We compare the average secondary network throughput (normalized by $M$) and the primary network throughput (normalized by $N$) assuming average PU-to-PU SNR=10dB over $T=20$ time slots, as a function of $p_{\text{MD,Target}}$ in Fig. 3(a) and Fig. 3(b), respectively, for four sensing policies. The first two policies employ fixed threshold selection in (3,4): the conventional myopic sensing policy with the bandwidth reward in (10) (myopic, imperfect) [10] and the myopic sensing policy that adapts to SU link SNR with the reward (11) (SU-SU CSI-aided, imperfect) [11]. The other two policies employ adaptive threshold selection (2,5) and the reward (9), where $R_{\text{FT}}^{mn}(t)$ is given by (10) for the myopic PU-SU CSI aided policy and (11) for the combined PU-SU and SU-SU CSI-aided myopic sensing policy. Moreover, the throughputs of the conventional and SU-SU CSI-adaptive strategies under perfect sensing are also plotted in the Fig. 3.

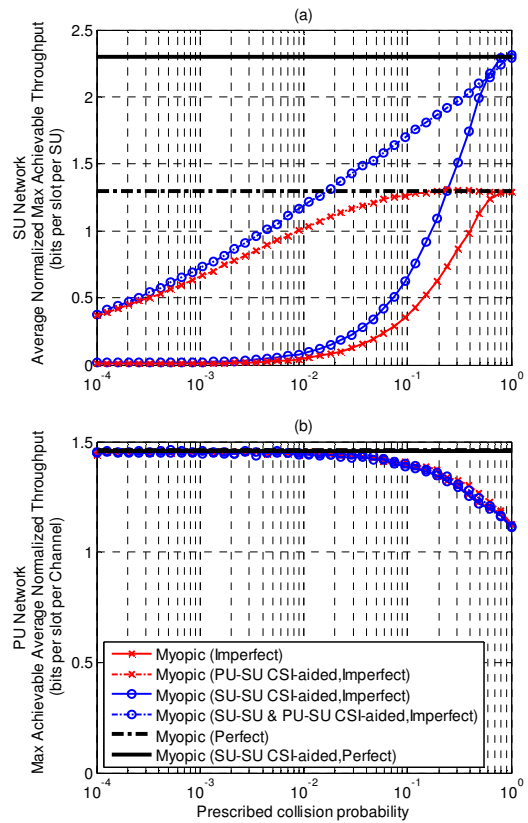

Fig. 3: Throughput vs. $p_{\text{MD,Target}}$ for (a) SU network; (b) PU network; 20 SU pairs; 40 channels; i.i.d Rayleigh fading; PU-to-SU SNR=SU-to-SU SNR $\bar{\gamma}=10dB$; PU-to-SU SNR $\bar{\lambda}=-10dB$; $T=20$; $v=100$. The legend for both is in (b).

Due to the miss detection rate constraint our proposed policy offers the same long-term protection to the PUs as conventional sensing strategies as demonstrated by overlapped PU performance curves in Fig. 3(b). The throughput of the PU network is compromised severely when $P_{\text{MD,Target}}>10^{-1}$ and approaches its optimal value as the prescribed collision probability tends to $10^{-2}$. However, from Fig. 3(a), the SU network throughput degrades rapidly for $P_{\text{MD,Target}} \leq 10^{-1}$ when conventional fixed threshold detection is employed. *The proposed threshold adaptation results in 0.4-1 bit per slot per SU throughput gain over the fixed threshold policy in the small* $P_{MD,Target}$ region. Both strategies converge to their ideal counterparts as the prescribed collision probability increases. *The combined adaptation provides up to 0.4 bits additional gain relative to adaptive sensing threshold selection alone for* $P_{MD,Target} \leq 10^{-1}$. Since both policies employ adaptive transmission, this gain is due to *adaptation to SU link CSI prior to sensing*.

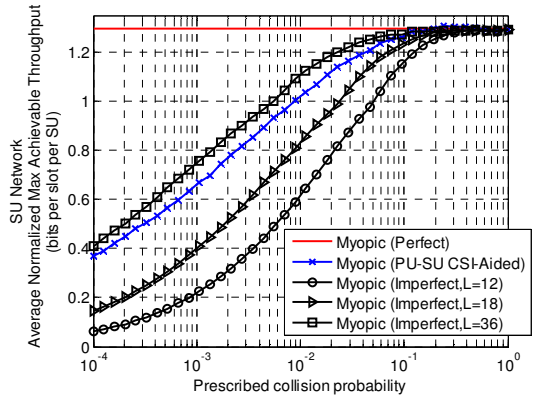

Fig. 4: Throughput of adaptive threshold selection and of cooperative sensing; myopic strategy; 20 SU pairs; 40 channels; i.i.d Rayleigh fading; $\bar{\gamma}=10dB$; $p_{MD,Target}=0.1$; $\bar{\lambda}=-10dB$; $T=20$; $\nu=100$.

However, in the low $p_{MD,Target}$ region adaptive threshold selection is more beneficial for the conventional myopic strategy than for the strategy that also adapts to the CSI of the SU link. First, the former strategy reaches the ideal sensor case for $p_{MD,Target}$ as small as 0.1 while the latter converges to the ideal case only for $p_{MD,Target}=1$. Moreover, at $p_{MD,Target}=10^{-2}$, the throughput gain provided by threshold adaptation is about 75% of the ideal throughput for the conventional myopic policy and is only 43% for the SU-to-SU CSI adaptive strategy. The lower relative gain in the latter strategy is due to reward adaptation that randomizes sensing decisions, so additional multiuser and multichannel diversity provided by sensing threshold adaptation has lower impact than for the conventional myopic strategy.

In Fig. 4 we evaluate the myopic policy using two spectrum detection approaches: sensing threshold adaptation and cooperative sensing. In the latter method, we assume OR-rule hard decision combining [4] where a fusion center collects independent individual sensing decisions from $L$ SUs and decides $H_1$ if any of the $L$ local decisions is $H_1$. The probability of miss detection and the probability of false alarm of the final decisions are $P_{MD}=p_{MD}^L$ and $P_{FA}=1-(1-p_{FA})^L$, respectively, where $p_{MD}$ and $p_{FA}$ are given by (3,4), and the threshold can be determined by inverting $P_{MD}$, i.e., $\tau = P_{MD}^{-1}(p_{MD,Target})$.

Cooperative sensing has lower throughput than the proposed PU-to-SU CSI-aided myopic policy unless the number of diversity branches is very large. We found that at least $L=30$ independent sensing observations are required to match the throughput of adaptive threshold selection at a single SU detector. Thus, *throughput improvement and multiuser*

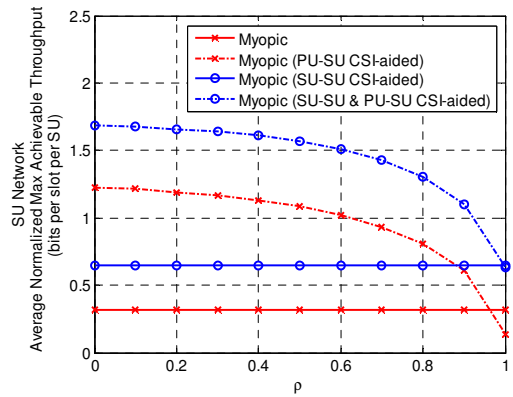

Fig. 5: Throughput vs. spatial correlation $\rho$; 20 SU pairs; 40 channels; log-normal fading; $\mu_{\gamma_{dB}}=10dB$; $\mu_{\lambda_{dB}}=-10dB$; $\sigma_{\gamma_{dB}}=\sigma_{\lambda_{dB}}=5dB$; $p_{MD,Target}=0.1$; $T=20$; $\nu=100$.

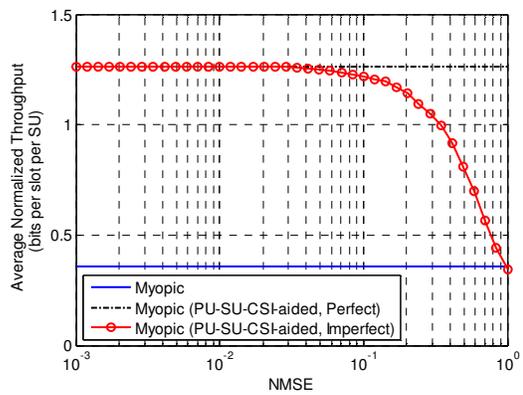

Fig. 6: Throughput vs. NMSE of CSI estimation; 20 SU pairs; 40 channels; i.i.d Rayleigh fading; $\bar{\gamma}=10dB$; $\bar{\lambda}=-10dB$; $p_{MD,Target}=0.1$; $T=20$; $\nu=100$.

*diversity gain of the proposed method outweigh the benefits of cooperative sensing for realistic CR networks.*

### B. Impact of Correlated Shadow Fading and CSI Error

We explore adaptation to the log-normal shadow fading where the short-term (multipath) fading is removed using diversity techniques. While estimation and tracking of shadow fading CSI is simpler and more practical than for short-term fading CSI for high speeds, the shadow fading signals from the PU transmitter to different SU sensors are likely to be correlated. We employ the correlated lognormal shadowing model [17] for the network with one PU transmitter and $M=20$ equally spaced SU detectors placed on a linear track[1]. The shadow fading coefficients are assumed uncorrelated across different channels and for all SU-to-SU links. For each channel, the correlation coefficient between any two PU-to-SU links observed at detectors $m$ and $m'$ is given by $\rho_{mm'}=\rho^{|m-m'|}$, where $\rho$ is the shadow fading correlation at two adjacent detectors. Each channel is modeled using the lognormal distribution with average dB-scale SNR $\mu_{\gamma_{dB}} = 10 dB$, $\mu_{\lambda_{dB}} = -10 dB$, and the dB-spread $\sigma_{\gamma_{dB}} = \sigma_{\lambda_{dB}} = 5 dB$. The impact of different values of $\rho$ is shown in Fig. 5. Note that the throughput of the proposed PU-to-SU CSI-aided sensing strategy degrades as the correlation $\rho$ increases although significant multiuser diversity gain is observed even for relatively high values of $\rho$. These results show that *the proposed method is useful in practical shadow fading scenarios* [17, 18].

---

[1] We assume that SU transmitter is responsible for spectrum sensing. In practice, sensing can also be carried out at the receiver side or at both ends of the SU link (equivalent to cooperative sensing with $L=2$).

As discussed in the introduction, estimated PU-to-SU channel gain will result in CSI mismatch, and CSI estimation errors can also degrade performance of proposed sensing threshold adaptation. We assume that the detector employs the Minimum Mean Square Error (MMSE) estimate of the actual PU-to-SU SNR $\lambda$ conditioned on its mismatched observation $\hat{\lambda}$. (We omit the indexes $m$, $n$ and $t$ for simplicity.) The threshold is calculated using the expected miss detection rate,

$$\hat{p}_{MD}(t) = \int_0^{+\infty} p_{MD}(t) f(\lambda | \hat{\lambda}) d\lambda, \quad (12)$$

$$\hat{\tau}(t) = \hat{p}_{MD}^{-1}(p_{MD,Target}), \quad (13)$$

where $p_{MD}(t)$ is given by (1) and $f(\lambda | \hat{\lambda})$ is the conditional probability density function (pdf) of $\lambda$ given $\hat{\lambda}$, e.g. [19]. The false alarm probability is computed using the threshold (13), and the reward is computed using (9) where $R_{FT}^{mn}(t)$ is given by (10). We illustrate the throughput vs. normalized mean-square-error (NMSE) of SNR estimation for the myopic strategy with adaptive threshold selection in Fig. 6.

We observe that the proposed approach approximates the ideal PU-to-SU CSI case when NMSE ≤ 0.1 and degrades gracefully to the conventional myopic policy with fixed threshold when the PU-to-SU CSI becomes unreliable. Note that NMSE ≥ 0.1 corresponds to severely degraded CSI prediction accuracy in conventional communication systems [20]. Thus, we conclude that *the proposed scheme is robust to PU-to-SU CSI mismatch.*

V. CONCLUSION

Adaptation of the detection threshold to the instantaneous SNR of the PU signal was proposed for CR spectrum sensing. The instantaneous miss detection probability constraint was imposed, and the resulting time-variant false alarm probability was incorporated into the sensing strategy design. It was demonstrated that the proposed sensing strategy randomizes sensing decisions and provides 0.4-1 bit per slot per SU throughput gain over the fixed threshold policy for small prescribed collision probabilities with the PU network and low average PU-to-SU SNR. Additional 0.4 bits can be gained by combined adaptation to PU-to-SU and SU-to-SU CSI. Moreover, cooperative sensing with at least 30 independent sensing results is necessary to match the throughput of proposed threshold adaptation at a single detector. Finally, it is shown that the proposed adaptive strategy is robust to shadow fading correlation and to CSI mismatch for practical CR network parameters.